\documentstyle[12pt, a4]{article}
\begin{document}
\author{B\"{u}lent G\"{o}n\"{u}l\footnote{gonul@gantep.edu.tr}~,
Okan \"{O}zer, Be\c{s}ire G\"{o}n\"{u}l and Fatma
\"{U}zg\"{u}n\and Department of Engineering Physics,Faculty of
Engineering,\and University of Gaziantep, 27310
Gaziantep-T\"{u}rkiye}
\title{Exact solutions of effective-mass Schr\"{o}dinger
equations}
\maketitle
\begin{abstract}
We outline a general method for obtaining exact solutions of
Schr\"{o}dinger equations with a position dependent effective mass
and compare the results with those obtained within the frame of
supersymmetric quantum theory. We observe that the distinct
effective mass Hamiltonians proposed in the literature in fact
describe exactly equivalent systems having identical spectra and
wave functions as far as exact solvability is concerned. This
observation clarifies the Hamiltonian dependence of the
band-offset ratio for quantum wells.
\end{abstract}


\section{Introduction}
An interesting question arises when one tries to formulate the
correct Hamiltonian for a particle with spatially dependent mass
in an arbitrary potential well. This problem often arises in the
calculation of confined energy states for carriers in
semiconductor quantum well structures under the envelope-function
and the effective-mass approximations where the effective mass of
a carrier is spatially dependent on the graded composition of the
semiconductor alloys used in the barrier and the well region of
the nanostructures. Since the mass and the momentum operators no
longer commute, the correct ordering of these operators within the
kinetic energy operator cannot be trivially assigned. This problem
of ordering ambiguity is a long standing one in quantum mechanics,
see for instance the excellent critical review by Shewell
\cite{shewell161959}. The study of quantum systems with position
dependent effective masses has been the subject of much activity
in recent years. Apart from being an interesting topic itself such
equations have found wide applications in the search of not only
electronic properties of semiconductors \cite{bastard1992} but
also quantum dots \cite{serra6671997}, quantum liquids
\cite{arias42481994}, $^{3}He$ clusters \cite{barranco89971997},
and metal clusters \cite{puente2831994}.

Although the treatment of effective-mass Schr\"{o}dinger equations
with non-constant mass is difficult, some exactly solvable models
for such systems have been recently introduced [7-11]. However,
from the physics point of view, none of these works seem
completely convincing. For example, Dekar and his co-workers
\cite{dekar1071999} solved exactly the one-dimensional
Schr\"{o}dinger equation, derived from the general form of the
effective-mass Hamiltonian without any restriction in the ordering
parameters, for a system with smooth potential and mass step. The
effective mass Hamiltonian and the connection rules for a system
with abrupt heterojunction were deduced from the study of the
limiting case when the smooth step potential and mass tend to an
abrupt potential and mass step, which suggest that the appropriate
form of the effective-mass Hamiltonian for an abrupt
heterojunction is the form proposed by \cite{bendaniel6831966}.
However, this modelling of the position dependent potential and
mass by a discontinuous profile is not physically acceptable,
because in the real world neither the potential nor the effective
mass can change abruptly across the heterojunction. In a different
perspective, Dutra and Almeida \cite{desouza252000} discussed the
relationship between exact solvability of the effective-mass
Schr\"{o}dinger equations and the ordering ambiguity. They
introduced a simple technique to transform mass dependent
Schr\"{o}dinger equations to the conventional Schr\"{o}dinger
equation with a constant mass to obtain in a closed form of the
full spectrum and wavefunctions of the original system. As an
illustrative example they considered a physically plausible case
(an exponentially changing mass and potential) and obtained the
corresponding exact solutions by mapping the system onto the well
known harmonic oscillator with centripetal barrier. It was nicely
clarified that in this case the energy levels are redefined in
such a way that the ordering ambiguity disappear. However their
model has a drawback, which works only for a restricted choice of
mass and potential functions to be able to serve exact solutions.
Although the works in Refs. \cite{plastino43981999,
milanovic70011999} within the frame of supersymmetric quantum
theory, together with a more recent work \cite{broy39612002} used
a Lie algebra, provide a powerful treatment for the systems of
interest, all the formulations introduced  in these models are
only valid for the BenDaniel and Duke Hamiltonian
\cite{bendaniel6831966} as they followed the work of Levy-Leblond
\cite{levy18451995}.

Notwithstanding, taking into account the spatial variation of
semiconductor type, some effective Hamiltonians have been proposed
with a coordinate-dependent mass for the carrier
[12,14-18], and many authors have been trying to determine the correct
Hamiltonian phenomenologically, for a review see
\cite{li127601993}. Although, there has been growing consensus in
favor of the BenDaniel-Duke Hamiltonian, see the related
references in \cite{li127601993}, the form of the effective-mass
Hamiltonian has been still a controversial subject due to the
location dependence of the effective mass and the question of the
exact form of the kinetic energy operator is still an open
problem. In the light of all these, the present work addresses a
more general and compact framework introducing two different
models to solve exactly for a physical system involving a position
dependent mass with the consideration of all physically acceptable
Hamiltonians proposed in the literature. We will make clear that
though the effective Hamiltonians appear in a different form, they
describe in fact exactly equivalent systems having identical
spectra and wavefunctions as far as exact solvability is
concerned. Although the literature covered similar problems, to
our knowledge an investigation such as the one we have discussed
in this paper was missing.

The paper is organized as follows. Using the spirit of the works
\cite{deL8431992} we introduce, in the next section, a generalized
exact treatment procedure for effective-mass Hamiltonians and use
the works in \cite{plastino43981999,milanovic70011999} to extend
their scenario and involve all possible Hamiltonians. Section 3
involves applications of the models developed for two different
but physically meaningful mass considerations. Discussion and
analysis of the results obtained are given in section 4. Finally,
some conlusions are drawn in section 5.

\section{Theoretical considerations}
The single-band effective-mass approach to a quantum-well problem
requires that the envelope function $\psi $ satisfy the
effective-mass equation
\begin{equation}\label{Hpsi}
\hat{H} \psi =E\psi
\end{equation}
where $E$ is the energy eigenvalue and $\hat{H}$ is the
Hamiltonian operator consisting of kinetic energy operator (
$\hat{T} $ ) and the potential energy operator ($\hat{V}$),
\begin{equation}\label{Hoperator}
\hat{H} =\hat{T}(\hat{z})+\hat{V}(\hat{z})
\end{equation}
Due to the compositional variation in a quantum well as a function
of location, the kinetic energy and the potential energy are
expressed as position-dependent operators in Eq.
(\ref{Hoperator}). The kinetic energy operator can be considered
to be composed of four elements: $1/\sqrt{m(\hat{z} )} $,
$1/\sqrt{m(\hat{z} )} $, $\hat{p} $, and $\hat{p} $, where
$\hat{z} $ and $\hat{p} $ are position and momentum operators,
respectively. Because $1/\sqrt{m(\hat{z} )} $ and $\hat{p} $  are
not commutable, there are different possible permutations to
represent the kinetic energy operator
[12,14-18]. All of these single-band effective-mass Hamiltonians are special
cases of a general form of the Hamiltonian introduced by von Roos
\cite{von75471981},
\begin{equation}\label{HvonRoos}
H_{vR} =\frac{1}{4} \left[ m^{\alpha }(\hat{z} ) \hat{p} m^{\beta
} (\hat{z} ) \hat{p} m^{\gamma } (\hat{z} ) +m ^{\gamma } (\hat{z}
) \hat{p} m ^{\beta }(\hat{z} ) \hat{p} m^{\alpha }(\hat{z} )
\right] +V(\hat{z} )~,
\end{equation}
where $\alpha +\beta +\gamma =-1$.

By the correspondence in wave mechanics $\hat{p} \rightarrow
-i\hbar d/dz $ and $\hat{z} \rightarrow z$, the effective mass
equation, Eq. (\ref{Hpsi}), togeher with any possible Hamiltonian
proposed in the literature as the special cases of Eq.
(\ref{HvonRoos}) can be written in a differential form,
\begin{equation}\label{schrwithmz}
-\frac{\hbar ^{2} }{2} \frac{d}{dz} \left[ \frac{1}{m(z)}
\frac{d\psi (z)}{dz} \right] +V^{eff} (z)\psi (z)=E\psi (z)~,
\end{equation}
where $V^{eff} (z)$ is termed the effective potential energy whose
algebraic form depends on the Hamiltonian employed (see Table
\ref{TABLE I}),
\begin{eqnarray}
V^{eff}(z)=V(z)+U_{\alpha \gamma}(z)
~~~~~~~~~~~~~~~~~~~~~~~~~~~~~~~~~~~~~~~~~~~~~~~~~~~~~~~~~~~~~~~~~~
\nonumber
\end{eqnarray}
\begin{equation}\label{effvz}
=V(z)-\frac{\hbar ^{2}}{4m^{3}(z)}\left[\left(\alpha
+\gamma\right)m(z)m''(z)-2\left( \alpha \gamma +\alpha +\gamma
\right) m'^{2} (z)\right],
\end{equation}
in which the first and second derivatives of $m(z)$ with respect
to $z$  are denoted by $m'$ and $m''$, respectively. The effective
potential is the sum of the real potential profile $V(z)$ and the
modification $U_{\alpha \gamma } (z)$  emerged from the location
dependence of the effective mass. A different Hamiltonian leads to
a different modification term.

In the following section we introduce two different theoretical
models to solve Eq. (\ref{schrwithmz}) exactly, which lead us to
define analytically solvable potentials for the systems
undertaken.

\subsection{Coordinate transformation method}
In a recent work, Roy and Roy \cite{deL8431992} have presented a
coordinate transformation method to solve analytically an
effective-mass Hamiltonian corresponding to the one proposed by
BenDaniel-Duke \cite{bendaniel6831966}. We extend and generalize
their model by tackling the problem with a fundamental point of
view, ie., without using a particular form of the effective
potential in (\ref{effvz}) and arrive at a conceptually consistent
result for exactly solvable systems with location dependent
masses. Our generalization is based on the work of De {\it et all}
\cite{deL8431992} in which they gave explicit point canonical
transformations and interrelated many shape-invariant exactly
solvable potentials within the frame of the usual Schr\"{o}dinger
equation with a constant mass. We show that their simple approach,
which consists of mapping through canonical transformations of
coordinates, can also be used to interrelate two distinct systems
having constant and position-dependent masses.

As the aim is to solve Eq. (\ref{schrwithmz}) in a closed form to
obtain the corresponding full energy spectrum and wavefunctions
exactly for a solvable potential, one needs to transform Eq.
(\ref{schrwithmz}) into a workable frame. Proceeding with a
transformation of both the independent and dependent variables of
the form
\begin{equation}\label{variables}
z=f(\tilde{z} )\quad ,\quad \psi (z)=\nu (\tilde{z} )\tilde{\psi } (\tilde{z%
} )\quad ,\quad m(z)=m\left[ f(\tilde{z} )\right] =\tilde{m}
(\tilde{z} )~,
\end{equation}
we transform Eq. (\ref{schrwithmz}) to
\begin{eqnarray}
-\frac{\hbar^{2}}{2\tilde{m}}\left\{\frac{d^{2}\tilde{\psi}}
{d\tilde{z}^{2}}+\left(\frac{2\nu'}{\nu}-\frac{f''}{f'}
-\frac{\tilde{m}'}{\tilde{m}}\right)\frac{d\tilde{\psi}}{d\tilde{z}}
-\left[\left(\frac{\nu'}{\nu}\right)\left(\frac{\tilde{m}'}{\tilde{m}}
+\frac{f''}{f'}\right)-\frac{\nu''}{\nu}\right]\tilde{\psi}\right\}~~~~~~~~~~
\nonumber
\end{eqnarray}
\begin{equation}\label{schrtransformed}
~~~~~~~~~~~~~~~~~~~~~~~~~~~~~~~~~~~~~~~~~~~~~~~~
+f'^{2}\left\{V^{eff}\left[f(\tilde{z})\right]-E\right\}
\tilde{\psi} =0~,
\end{equation}
in which the prime denotes differentiation with respect to
$\tilde{z} $ . One can now easily reduce Eq.
(\ref{schrtransformed}) to the Schr\"{o}dinger equation with
constant mass, which requires
\begin{equation}\label{vzfz}
\nu (\tilde{z} )=C\sqrt{f'(\tilde{z} )\tilde{m} (\tilde{z}
)}~,~f'(\tilde{z} )=\sqrt{\frac{m_{0} }{\tilde{m} (\tilde{z} )}
}~,
\end{equation}
where $m_{0} $ stands for a constant effective mass, and $C$  is a
constant of integration. Eq. (\ref{schrtransformed}) in this case
reads
\begin{equation}\label{schrmbar}
-\frac{\hbar^{2}}{2m_{0}}\frac{d^{2}\tilde{\psi}}{d\tilde{z}^{2}}
+\left\{\frac{\hbar^{2}}{2m_{0}}\left(\frac{5}{16}\frac{\tilde{m}'^{2}
}{\tilde{m}^{2}}-\frac{1}{4}\frac{\tilde{m}''}{\tilde{m}}\right)
+V^{eff}\left[f(\tilde{z})\right]\right\}\tilde{\psi}(\tilde{z})
=E\tilde{\psi}(\tilde{z})~.
\end{equation}
It is important to note that Eqs. (\ref{schrwithmz}) and
(\ref{schrmbar}) have identical spectra. With the wavefunctions
$\psi (z)$ being square integrable, ie., $\left\langle \psi
(z)\right.\left| \psi (z)\right\rangle =1 $ , setting the integral
constant $C=1/\sqrt{m_{0} } $  in Eqs. (\ref{variables}) and
(\ref{vzfz}) and having in mind that $m(\tilde{z})>0$ we find
that $\left\langle \tilde{\psi } (\tilde{z} )\right.\left| \psi
(\tilde{z} )\right\rangle =1$ as well, ie., the functions
$\tilde{\psi } (\tilde{z} )$ are also square integrable.

For further simplification, we recall the mathematical definition
used in arriving at Eq. (\ref{vzfz}), $f'^{2} (\tilde{z} )\left[
\tilde{m} (\tilde{z} )/m_{0} \right] =1$, and remind that it may
be inverted and written as
\begin{equation}\label{zbar}
\tilde{z} =\frac{1}{\sqrt{m_{0} } }
\int\limits_{0}^{z}\sqrt{m(z)}~dz=f^{-1}(z)~,
\end{equation}
which defines (though in implicit form) the mapping function
$f(\tilde{z} )$ and consequently enables finding $\tilde{m}
(\tilde{z} )$ , together with the effective potential $V^{eff}
\left[ f(\tilde{z} )\right] $. Bearing in mind that
\begin{equation}\label{mbarprime}
\tilde{m}'=\frac{d\tilde{m}(\tilde{z})}{d\tilde{z}}=
\frac{dz}{d\tilde{z}}\frac{d}{dz}\left[\tilde{m}(\tilde{z})\right]
=f'(z)m'(z) =\sqrt{\frac{m_{0}}{m(z)}}~\frac{dm(z)}{dz}~,
\end{equation}
Eq. (\ref{schrmbar}) can also be reduced to the form,
\begin{equation}\label{schrnormal}
-\frac{\hbar ^{2} }{2m_{0} } \frac{d^{2}\tilde{ \psi } (\tilde{z}
)}{d\tilde{z} ^{2} } +\left\{ \frac{\hbar ^{2} }{32m^{3} (z)}
\left[ 7m'^{2} (z)-4m(z)m''(z)\right] 
+ V^{eff} (z) \right\} _{z=\tilde{z} }
\tilde{\psi } (\tilde{z} ) = E\tilde{\psi } (\tilde{z} )
\end{equation}
It is worth to note that the change of variable in (\ref{zbar})
may not always be invertible or not easily be invertible. But this
does not really pose a problem as far as solvability of Eq.
(\ref{schrwithmz}) is concerned. This is because the new
coordinate $\tilde{z} $  as a function of the old coordinate $z$
is explicitly known from (\ref{zbar}) and if we choose $V^{eff} $
such that
\begin{equation}\label{veffpotinz}
V^{eff} (z)=V_{ES} (\tilde{z} )-V_{m} (\tilde{z} )~,
\end{equation}
where $V_{ES} $ is an exactly solvable potential and $V_{m} $  is
the mass-dependent part of the full potential,
\begin{equation}\label{vmpot}
V_{m} =\frac{\hbar ^{2} }{32m^{3} (z)} \left[ 7m'^{2}
(z)-4m(z)m''(z)\right]_{z=\tilde{z}}~,
\end{equation}
obviously one then finds exactly the spectrum of the system in
(\ref{schrwithmz}). The corresponding wavefunctions can be
obtained using Eq. (\ref{variables}).

The interest in exactly solvable problems in quantum physics has
increased sharply in the last few years. This is concerned of
course with the fact that the behavior description of some
physical systems is usually very complicated, but in some cases
such systems can be modelled by means of a quite simple
Hamiltonian which leads to standard problems of quantum mechanics,
as the one presented in this section. As the spectral properties
of the constant-mass Schr\"{o}dinger equation of solvable
potentials are well known in the literature, one can readily
obtain a corresponding potential for the effective-mass
Schr\"{o}dinger equation with identical spectra. The problem of
generating isospectral potentials in quantum mechanics has been
considered for more than 50 years, but recently the research
efforts on this topic have been considerably intensified.

In the next section, we discuss an alternative treatment for the
problem under consideration and illustrative examples will be
given in section 3.

\subsection{Supersymmetric approach}
In the light of the recent works \cite{plastino43981999,
milanovic70011999}, we introduce a superpotential $W(z)$ and the
associated pair of operators $A$ and $A^{+} $ defined by
\begin{equation}\label{operators}
A\psi (z)=\frac{\hbar }{ \sqrt{2m(z)} } \frac{d\psi (z)}{dz}
+W(z)\psi (z)~,~A^{+} \psi (z)=-\frac{d}{dz} \left[ \frac{\hbar
\psi (z)}{\sqrt{2m(z)} } \right] +W(z)\psi (z)
\end{equation}
Notice that, due to the location dependence of the mass, $d/dz$
and $\hbar/\sqrt{2m(z)}$  do not commute anymore. Within the frame
of supersymmetric quantum mechanics \cite{cooper2671995}, the
partner Hamiltonians read
\begin{eqnarray}
H_{1} =A^{+} A=-\frac{\hbar ^{2} }{2m(z)} \frac{d^{2} }{dz^{2} }
-\left( \frac{\hbar ^{2} }{2m(z)} \right)'\frac{d}{dz} +\left[
W^{2} (z)-\left( \frac{\hbar W(z)}{\sqrt{2m(z)} } \right)'
\right]~,
\nonumber
\end{eqnarray}
\begin{equation}\label{h1andh2}
H_{2} =AA^{+} =H_{1}
+\frac{2\hbar W'(z)}{\sqrt{2m(z)} } -\frac{\hbar }{\sqrt{2m(z)} }
\left( \frac{\hbar }{\sqrt{2m(z)} } \right)''~.
\end{equation}

We see that the two partner Hamiltonians describe particles with
the same efective mass-spatial dependence but in different
potentials. As the usual exactly solvable potentials, as well as
many recently discovered ones, are given in the form of the first
partner potential in (\ref{h1andh2}), we will work through the
applications in the next section with the consideration that the
effective potential in (\ref{effvz}) should satisfy the condition,
\begin{equation}\label{effpotinz}
V^{eff} (z)=V_{1}=W^{2} (z)-\left( \frac{\hbar W(z) }{\sqrt{2m(z)}
} \right)'~.
\end{equation}
The second partner potential is used for obtaining the
corresponding energy spectrum. Considering the shape invariance
concept \cite{cooper2671995}, the supersymmetric partner
potentials in (\ref{h1andh2}) obey the integrability condition
\begin{equation}\label{susypotdefinition}
V_{2} (z,a_{1})=V_{1} (z,a_{2})+R(a_{1})~,~a_{2}=f(a_{1})~.
\end{equation}
The above relation means that the partner potentials $V_{1}$  and
$V_{2}$ have the same form, but are characterized by different
values of the parameter sets. The energy eigenvalues, $E_{n}
=\sum\limits_{i=1}^{n}R(a_{i} )$, and eigenfunctions of shape
invariant poteantials can be obtained in algebraic fashion. For
more details the reader is refered to \cite{plastino43981999,
milanovic70011999, cooper2671995}.

\section{Applications}
A square quantum well subject no external field is usually modeled
by a step profile which discontinues at the heterojunctions. The
commonly used interface conditions at the heterojunctions are: (i)
the continuity of the envelope function and its first derivative,
or (ii) the continuity of the envelope function and its first
derivative divided by the effective mass. Within each flat region
of the square quantum well, the mass is a constant, and the
effective potentials, Eq. (\ref{effvz}), are identical because the
derivatives of the mass with respect to the position vanish. Thus
envelope functions within each flat region are independent of the
Hamiltonian used for the analysis. If the same interface
conditions at the discontinue points between two adjacent regions
are essentially imposed for the distinct effective Hamiltonians,
the eigensolutions will be exactly identical. As a result, the
effective-mass Hamiltonians in (\ref{effvz}) are expected  to
produce the same transition energy, namely, the band-offset ratio
will be independent of the Hamiltonian if the heterojunction is
modeled by a step function with essentially imposed interface
conditions. This point will be discussed in the next section
within the frame of our results.

However, the discontinuity of the square quantum-well model
implies an infinite external electric field at the
heterojunctions, and this is not physically possible. In reality,
the potential changes over a few monolayers for a perfect
microscopic interface. In this section, the square quantum well is
modeled by more realistic smoothed profiles, which leads to
different effective potential descriptions. This modified
realistic potential profiles remove the discontinuity of the sharp
square quantum well. Considering the realistic quantum well
applications in the works \cite{li127601993, datta1989}, we test
our models introduced in section 2 employing the two different but
physically meaningful effective mass variations discussed in the
following.

\subsection{For $m(z)=m_{0}\exp(\lambda z)$}
To demonstrate the simplicity of the models used we consider here
a particular case, as a first illustrative example, for which one
have exact solutions. That is a particle with exponentially
increasing or decaying, $|\lambda|\propto 1/L$ with $L$ being the
quantum-well width, in the presence of a potential with similar
behaviour that will be defined for each Hamiltonian in this
section.

Starting with Eqs. (\ref{schrnormal}-\ref{vmpot}), within the
framework of the coordinate transformation technique
and taking the harmonic oscillator potential as a
solvable potential, $V_{ES} (z)=Bz^{2} $ , and bearing in mind
that $\tilde{z} =\left( 2/\lambda \right) \exp \left( \lambda z/2
\right) $  from Eq. (\ref{zbar}), we obtain a variety of exactly
solvable effective potential descriptions appearing in the
original effective mass equation, Eq. (\ref{schrwithmz}). These
are, corresponding to distinct effective mass Hamiltonians
presented by Eqs. (\ref{HvonRoos}) and (\ref{effvz}),
\begin{equation}\label{veffbdd}
V_{BDD}^{eff} (z)=V_{0} \exp (\lambda z)-\frac{3\hbar ^{2} \lambda ^{2} }{%
32m_{0} } \exp (-\lambda z)\quad ,\quad \alpha =\gamma =0\quad
,\quad \beta =-1~,
\end{equation}
for the BenDaniel-Duke Hamiltonian \cite{bendaniel6831966},
\begin{equation}\label{veffgw}
V_{G-W}^{eff} (z)=V_{0} \exp (\lambda z)+\frac{5\hbar ^{2} \lambda ^{2} }{%
32m_{0} } \exp (-\lambda z)\quad ,\quad \beta =\gamma =0\quad
,\quad \alpha =-1~,
\end{equation}
for the Gora and Williams (or the Bastard ) Hamiltonian
\cite{gora11791969, bastard56931981}, and
\begin{equation}\label{veffzk}
V_{Z-K}^{eff} (z)=V_{0} \exp (\lambda z)+\frac{\hbar ^{2} \lambda ^{2} }{%
32m_{0} } \exp (-\lambda z)\quad ,\quad \alpha =\gamma
=-\frac{1}{2} \quad ,\quad \beta =0~,
\end{equation}
for the Zhu and Kroemer Hamiltonian \cite{zhu35191983}. We also
obtain exactly the same result as in (\ref{veffzk}) for the
Li-Kuhn Hamiltonian \cite{li127601993}
($\beta=\gamma=-1/2\;,\alpha=0$). In the above equations,
$V_{0}=4B/\lambda^{2}=\hbar^{2}\lambda^{2}/32m_{0}$. From Eqs.
(\ref{schrwithmz}) and (\ref{schrnormal}), it is clear that though
the appearance and behaviour of the potentials in
(\ref{veffbdd})-(\ref{veffzk}) are dissimilar, they have identical
spectra, $E_{n}=(n+1/2)\hbar\omega$, where
$\omega=\sqrt{2B/m_{0}}$. This result supports the similar
discussion presented in the section 3 of a recently published
paper \cite{broy39612002}. Moreover, from Eqs. (\ref{variables})
and (\ref{vzfz}), these Hamiltonians have the same wavefunctions,
$\psi _{n}(z)=N_{n}\left[\exp(\lambda
z)\right]^{1/4}\tilde{\psi}_{n}(\tilde{z})$ where
$\tilde{\psi}_{n}(\tilde{z})=\left\{\frac{1}{2^{n}n!}
\sqrt{\frac{\eta}{\pi}}\right\}^{1/2}H_{n}\left(\sqrt{\eta}\tilde{z}\right)
\exp\left(-\frac{\eta\tilde{z}^{2}}{2}\right)$ with
$\eta=\sqrt{2Bm_{0}}/\hbar$. As a result, the distinct
effective Hamiltonians considered here are not only isospectral
but also describe identically equivalent systems as far as exact
solvability is concerned. This significant result is also
confirmed below with use of the supersymmetric approach described
in section 3 as an alternative treatment.

For the supersymmetric considerations, in the light of Eq. (35) of
\cite{plastino43981999} we propose a superpotential
\begin{equation}\label{Wforexpx}
W(z)=\frac{\sqrt{2m_{0} } \delta }{\hbar \lambda } \exp (\lambda z/2) -%
\frac{\hbar \lambda }{4\sqrt{2m_{0} } } \exp (-\lambda z/2)~,
\end{equation}
where $\delta =\hbar \omega $ , which yields the supersymmetric
partner potentials in (\ref{h1andh2}) in the form
\begin{eqnarray}
V_{1} (z)=V_{0} \exp (\lambda z)-\frac{3\hbar ^{2} \lambda ^{2}
}{32m_{0} } \exp (-\lambda z)-\frac{\delta }{2}~, \nonumber
\end{eqnarray}
\begin{equation}\label{Viinzcoordinate}
V_{2} (z)=V_{0} \exp (\lambda z)-\frac{3\hbar ^{2} \lambda ^{2}
}{32m_{0} } \exp (-\lambda z)+\frac{\delta }{2}~,~~~~
\end{equation}
which is the simplest case of the shape invariance integrability
condition given by (\ref{susypotdefinition}) due to the partner
potentials in (\ref{Viinzcoordinate}) differing only by a uniform
energy shift by $\delta $ . Note that the first partner has the
same shape as in (\ref{veffbdd}) corresponding to the
BenDaniel-Duke effective potential. Considering the shifting term
$\delta /2 $ , together with Eq. (\ref{susypotdefinition}), one
can easily find the corresponding energy spectrum, $E_{n}
=(n+1/2)\hbar \omega $, which overlaps with the one found through
the transformation technique. For the other exactly solvable shape
invariant effective potentials we use a simple expression,
\begin{equation}\label{exacsoleffpot}
V_{ES}^{eff} (z)=V_{BDD}^{eff} (z)-U_{\alpha \gamma } (z)=V_{1}
(z)-U_{\alpha \gamma } (z)~,
\end{equation}
in which $U_{\alpha \gamma } (z)$  is the modification term in Eq.
(\ref{effvz}),
\begin{equation}\label{Uforexpx}
U_{\alpha \gamma }^{G-W} (z)=-\frac{\hbar ^{2} \lambda ^{2}
}{4m_{0} } \exp (-\lambda z)~~~,~~~U_{\alpha \gamma
}^{Z-K} (z)=U_{\alpha \gamma }^{L-K} =-\frac{\hbar ^{2} \lambda
^{2} }{8m_{0} } \exp (-\lambda z)~.
\end{equation}

The substitution of $V_{1} (z)$ in Eq. (\ref{Viinzcoordinate})
into Eq. (\ref{exacsoleffpot}) leads us to arrive at Eqs.
(\ref{veffgw}) and (\ref{veffzk}), which confirms the reliability
of the coordinate transformation technique developed in the
previous section. Once more it is clear that the potentials in
(\ref{veffbdd}-\ref{veffzk}) will have the same energy spectra and
wavefunctions as a unique superpotential is used, (Eq.
\ref{Wforexpx}), for the generation of these analytically solvable
potentials within the frame of supersymmetric quantum theory. For
the relation between the superpotential and wavefunction in case
of the supersymmetric considerations the reader is referred to
\cite{plastino43981999, milanovic70011999}.

\subsection{For $m(z)=m_{0} \left ( \frac{a + q^{2}}{1+q^{2}} \right)^{2}$}
This physically convenient mass variation, in which $a$ is a
positive constant and $q(=\bar\lambda z)$ involves the variable
with a positive width parameter $\bar\lambda \propto 1/L$,
considered here to convince the reader for that the models
introduced in this paper works for all smoothly varying masses,
unlike the recent models \cite{dekar1071999, desouza252000}. There
is another reason for the consideration of the work presented in
this section. Among various potential shapes (quantum well
profiles), there is one which has attracted some research
attention recently: this is the case of P\"{o}schl-Teller
potential with a constant electron mass \cite{haley2371997,
todorovic2682001} . Unfortunately, this idealized potential is not
realizable in the common, graded ternary alloy based quantum
wells, because of the effective mass therein necessarily varies,
together with the potential. In this respect, the result shown in
this section and its related discussion given in the next section
would be helpful in designing realistic ternary alloy based
structures with properties equivalent to those of idealized
P\"{o}schl-Teller potential.

From Eq. (\ref{zbar}), the relation between the transformed
coordinate and the old one is $\tilde{z}=z+(1/\bar\lambda
)(a-1)tan^{-1} q$, and considering the P\"{o}schl-Teller potential
as one possible choice from analytically solvable potential family
and working within the framework of the coordinate transformation
method, one arrives at a class of exactly solvable potentials for
the use in Eq. (\ref{schrwithmz}) belonging to different effective
mass Hamiltonians,
\begin{eqnarray}
V^{eff}_{BDD}(z)=V_{PT}(\tilde{z})
+\frac{(a-1)\left[3q^{4}+q^{2}(4-2a)-a\right]\hbar^{2}\bar\lambda
^{2}}{2m_{0}(q^{2}+a)^{4}}~,
\nonumber
\end{eqnarray}
\begin{eqnarray}
V^{eff}_{Z-K}(z)=V_{PT}(\tilde{z})
-\frac{(a-1)\left[3q^{4}+2q^{2}-a\right]\hbar^{2}\bar\lambda
^{2}}{2m_{0}(q^{2}+a)^{4}}~,
\nonumber
\end{eqnarray}
\begin{eqnarray}
V^{eff}_{L-K}(z)=V_{PT}(\tilde{z})
+\frac{(a-1)q^{2}\hbar^{2}\bar\lambda^{2}}{2m_{0}(q^{2}+a)^{4}}~,
\nonumber
\end{eqnarray}
\begin{equation}\label{potforqm}
V^{eff}_{BDD}(z)=V_{PT}(\tilde{z})
-\frac{(a-1)\left[3q^{4}+q^{2}(6-4a)-a\right]
\hbar^{2}\bar\lambda^{2}}{2m_{0}(q^{2}+a)^{4}}~,
\end{equation}
where
$V_{PT}(\tilde{z})=-A\left(A+\frac{\bar\lambda
\hbar}{\sqrt{2m_{0}}}\right)sech^{2}\left\{\bar\lambda
\left[z+\frac{(a-1)tan^{-1}~q}{\bar\lambda} \right]\right\}$.
Note that for $a\rightarrow 1~,m(z)\rightarrow m_{0},$ all the
above effective potentails reduce to the conventional
P\"{o}schl-Teller potential, $V^{eff}(z)\rightarrow V_{PT}(z)$,
due to $\tilde{z}\rightarrow z$. From Eqs. (\ref{schrwithmz}), (\ref{vzfz}) and
(\ref{schrnormal}), the bound state energy spectra and
wavefunctions corresponding to the potentials in (\ref{potforqm})
are $E_{n}=-\left(A-\frac{n\bar\lambda\hbar}{\sqrt{2m_{0}}}\right)$ and
$\psi_{n}(z)=N_{n}\sqrt{\frac{a+q^{2}}{1+q^{2}}}~\tilde\psi(\tilde
z )$ with $\tilde{\psi}(\tilde{z})$ being the well described wavefunction \cite{flugge1971}
for the solution of the usual Schr\"{o}dinger equation with a
constant-mass P\"{o}schl-Teller potential.

All these results are fully confirmed with  use of the
supersymmetric expressions presented in section 2 by adopting an
ansatz for the superpotential,
\begin{equation}\label{Wforqm}
W(z)=A~tanh\left\{\bar\lambda
\left[z+\frac{(a-1)tanh^{-1}~q}{\bar\lambda}\right]\right\}
+\frac{(a-1)~q~\hbar~\bar\lambda}{\sqrt{2m_{0}}(a+q^{2})^{2}}~.
\end{equation}

\section{Discussion}
We discuss here the physics behind the results obtained. First,
the applications given in the previous sections make clear the
band-offset ratio dependence on the effective mass Hamiltonians,
which is significant for quantum well applications. The
conduction-band-offset ratio, which is the ratio of the
conduction-band offset to the total band gap of the
heterojunction, has been investigated in various quantum wells
because of its fundamental importance and application. The ratio
has been measured by spectroscopic  and electrical methods. From
Duggan's and Kroemer's review articles \cite{duggan12241985} about
the experimental and theoretical works, it can be seen that
spectroscopic techniques are preferred over electrical ones in
exploring the band-offset ratio via a quantum well, and that
researchers using spectroscopy usually try to match their data
with the theoretical results to determine the band offset. To
demonstrate simply the band-offset ratio variation due to the
choice of the Hamiltonian, we will focus here on transition
energies between levels in conduction and valence bands. Following
the works \cite{li127601993, chomette38351986} and considering
only single-band effective-mass equations for the electron and the
hole, Eq. (\ref{schrwithmz}), one finds the transition energy in
the form of $E_{T}=E_{e}+E_{h}+E_{G}$ where $E_{e}, E_{h}$ are the
eigenenergies in the conduction and valence bands, respectively,
and $E_{G}$ is the band-gap energy. From the previous sections, it
is obvious that the effective Hamiltonians undertaken will yield
the same transition energies between identical transition levels
due to the identical values for $E_{e}$ and $E_{h}$. Hence, the
band-offset ratio for the BenDaniel-Duke Hamiltonian and for the
others can be found by solving
$E^{others}_{T}(Q^{others})=E^{BDD}_{T}(Q^{BDD})$, with $Q$ being
the band-offset ratio. Recall the relation between the conduction
band-offset ratio and the conduction-band-offset energy, $V_{cb}=Q
\Delta E_{g}$, where $\Delta E_{g}$ denotes the band-gap
difference between binary and ternary materials. Having in mind
that the band-offset ratio of a quantum well determines the
barrier height of the conduction band and valence bands, $V_{cb}$
corresponds to the depth of the effective potentials discussed
through the article. This leads to the realization of the
band-offset ratio dependence on the effective-mass Hamiltonian due
to the underlying differences between the strengths of the
potentials obtained. As a result of this, in the interpretation of
a given spectrum, the Hamiltonian employed in the analysis cannot
be regarded independently of the band-offset ratio utilized,
unlike the case of a simple square quantum well consideration.
Therefore, an attempt to determine the band-offset ratio from
experimental data by matching calculated transition energies with
spectral peaks would involve large inaccuracy.

There is another interesting point behind the present results. A
systematic procedure based on the inverse spectral theory and
supersymmetric transformations has been recently proposed
\cite{todorovic2682001} for optimized design of semiconductor
quantum well structures via tailoring the quantum well potential,
which enables shifting bound states in a quantum well and makes
the search for the best desired energy spectrum and potential
shape. By varying the free parameters appearing in the procedure
one can then design a convenient optimized structure. However, in
these notable works the idealized constant-mass P\"{o}schl-Teller
potential, which allows one to set analytically the spacing
between states, considered since a direct implementation of a more
realistic position-dependent effective mass related to the
position-dependent potential in their theoretical considerations
is not trivial. The optimization of continuously graded structures
thus require more sophisticated techniques.  In this respect, we
believe that the applications given in the previos section, in
particular the one involved the P\"{o}schl-Teller potential, give
a lot of material for experimenting in the optimized quantum well
laser design. For instance, though we have explicitly shown that
the energy spectra of the realistic ternary alloy based structures
with a carrier having a spatially varying mass are equivalent to
those of the constant-mass  P\"{o}schl-Teller Potential, the
maximization of the gain may be accomplished via changing the
quantum well profile which is in turn changes the wave functions.
Hence the consideration of our results within supersymmetric
transformations, as an alternative to the recent procedure
proposed, in particular the significant difference between the
wavefunctions by $\nu(\tilde{z})$ shown in (\ref{variables}),
relating the solutions of the Schr\"{o}dinger equations with
constant and location dependent masses, and use of the more
realistic effective potentials in (\ref{potforqm}) in tailoring
process instead of the standard P\"{o}schl-Teller potential,
should reproduce more reliable results in order to help for  the
best design of such structures. Along this line the work is in
progress.

Moreover, Plastino and his co-workers \cite{plastino782000}
recently studied some simple one-dimensional quantum mechanical
systems characterized by a piecewice flat potential and mass to
illustrate the influence of a non-constant mass on the density of
the bound state energy levels. With the consideration of a finite
potential well they showed that the number of bound states is less
than those of the constant mass situation when the effective mass
inside the well is lower than that of outside ($m_{0}$), and the
opposite behaviour occurs when the effective mass inside the well
is larger than the mass outside. However, our applications in
section 3 do not confirm their work. This contradiction may raise
a further discussion on the reliability of the present results,
which can be clarified  as follows. In the two different variable mass
definitions, $m(z)$, for the quantum well used in the previous
section, $m(z)>m_{0}$ for the case $\lambda > 0$ and $a > 1$ while
$m(z)<m_{0}$ in case $\lambda < 0$ and $a < 1$. The
consideration of $m(z)>m_{0}$ case leads to single potential
wells, sharper than that of the standard potential corresponding
to constant-mass potentials, whereas $m(z)<m_{0}$ case give rise
to bistable double well potentials, like the related illustrations
in \cite{plastino43981999}. In spite of the different aspects
exhibited by the effective potentials defined in section 3, we
have clearly shown that they share the same energy spectra
regardless of $\lambda$ and $a$ values, together with the
constant-mass potentials ($\lambda=0, a=1$). Consequently, one
expects that the density of bound states for the systems we
interest should not depend on the variation of a carrier mass.

\section{Concluding remarks}
In this article, we have discussed the problem of solvability and
ordering ambiguity in quantum mechanics as the form of the
effective-mass Hamiltonian has been a controversial subject due to
the location dependence of the effective mass. It was shown
through particular examples that the exact solvability depends not
only on the form of the potential, but also on the spatial
dependence on the mass. Within the framework of the two different
theoretical treatments, the effective-mass Schr\"{o}dinger
equation has been transformed to a constant-mass Schr\"{o}dinger
equation and we have clarified that the Schr\"{o}dinger equations
with different masses and potentials can be exactly isospectral.
We have also shown that though the potential energy of the
BenDaniel-Duke Hamiltonian differs from  the effective potentials
of other Hamiltonians proposed in the literature by a term caused
by the mass dependence on location, the exact analytical solutions
to the effective-mass equations do not change with the
Hamiltonian. As far as we know, this feature was not perceived
until now. The discussion given behind the results obtained may be
of interest, e.g., in the design and optimization of semiconductor
quantum wells. In addition to their practical applications, we
believe that the study of quantum mechanical systems with a
position-dependent mass within the framework of the present models
will raise many interesting conceptual problems of fundamental
nature. In particular, the methods used in this article may be
extended to find applications in also the study of quasi- and
conditionally-exactly solvable systems with non-constant masses.

\newpage\
\begin{table}
\caption{\label{TABLE I}
        { Single-band effective mass Hamiltonians, Eq.
(\ref{HvonRoos}).
        }}
\vspace{5mm}
\begin{center}
\begin{tabular}{cccc}
\hline \hline
\\
Hamiltonian&$\alpha$&$\beta$&$\gamma$
\\
\hline
\\
Ref. \cite{bendaniel6831966}&0&-1&0\\ \\ Ref. \cite{gora11791969,
bastard56931981}&-1&0&0\\ \\ Ref. \cite{zhu35191983}&-1/2&0&-1/2\\
\\ Ref. \cite{li127601993}&0&-1/2&-1/2\\ \\ \hline
\end{tabular}
\end{center}
\end{table}

\newpage\

\end{document}